\documentclass[aps,prb,reprint,superscriptaddress]{revtex4-2}
\usepackage{graphicx}
\usepackage{dcolumn}
\usepackage{bm}
\usepackage{amsmath}
\usepackage{braket}
\usepackage{verbatim}
\usepackage{xcolor}

\begin{document}

\title{Magnetic and nematic order of Bose-Fermi mixtures in moir\'e superlattices of 2D semiconductors}

\author{Feng-Ren Fan}
\thanks{These two authors contributed equally}
\affiliation{New Cornerstone Science Laboratory, Department of Physics, The University of Hong Kong, Hong Kong, China}
\affiliation{HK Institute of Quantum Science and Technology, University of Hong Kong, Hong Kong, China}

\author{Tixuan Tan}
\thanks{These two authors contributed equally}
\affiliation{HK Institute of Quantum Science and Technology, University of Hong Kong, Hong Kong, China}
\affiliation{Department of Physics, Stanford University, Stanford, CA 94305, USA}

\author{Chengxin Xiao}
\affiliation{New Cornerstone Science Laboratory, Department of Physics, The University of Hong Kong, Hong Kong, China}
\affiliation{HK Institute of Quantum Science and Technology, University of Hong Kong, Hong Kong, China}

\author{Wang Yao}
\email{wangyao@hku.hk}
\affiliation{New Cornerstone Science Laboratory, Department of Physics, The University of Hong Kong, Hong Kong, China}
\affiliation{HK Institute of Quantum Science and Technology, University of Hong Kong, Hong Kong, China}

\begin{abstract}
We investigate the magnetic orders in a mixture of a boson (exciton) and a fermion (electron or hole) trapped in transition-metal dichalcogenides moir\'e superlattices. A sizable antiferromagnetic exchange interaction is found between a carrier and an interlayer exciton trapped at different high symmetry points of the moir\'e supercell. This interaction at a distance much shorter than the carrier-carrier separation  dominates the magnetic order in the Bose-Fermi mixture, where the carrier sublattice develops  ferromagnetism opposite to that in the exciton sublattice. We demonstrate the possibility of increasing the Curie temperature of moir\'e carriers through electrical tuning of the exciton density in the ground state. In a trilayer moir\'e system with a p-n-p type band alignment, the exciton-carrier interplay can establish a layered antiferromagnetism for holes confined in the two outer layers. We further reveal a spontaneous nematic order in the Bose-Fermi mixture, arising from the interference between the Coulomb interaction and p-wave interlayer tunneling dictated by the stacking registry.
\end{abstract}

\maketitle

\emph{\color{blue} Introduction.}
Moiré superlattices in the twisted layered structures of 2D crystals have sparked tremendous interest as highly tunable physical platforms that can host exotic correlated states of matter~\cite{cao2018unconventional,Andrei_2020, balents2020superconductivity, Kennes_2021, andrei2021marvels, wilson2021excitons,regan2022emerging,lau2022reproducibility}.
In particular, transition metal dichalcogenides (TMD) as a semiconducting building block have enabled the exploration of both fermionic (electronic) and bosonic (excitonic) many-body phenomena on a common moir\'e platform~\cite{wilson2021excitons,regan2022emerging,huang2022excitons}.The moir\'e of twisted TMDs forms superlattice confinement for electron and hole, as well as for exciton as their bound state~\cite{yu2017moire,seyler2019signatures,jin2019observation,tran2019evidence,alexeev2019resonantly,brotons2020spin}. 
At a heterointerface where hole energetically favors one layer while electron favors the other, excitons have an interlayer configuration which endows them an ultralong lifetime and a permanent electrical dipole~\cite{rivera2018interlayer}. These bosonic composite particles can be injected either optically, or electrically through electron hole co-doping~\cite{wang2019evidence}.
The strong dipole-dipole interaction between excitons trapped in the moir\'e lattice has enabled a playground for studying Bose-Hubbard type physics~\cite{li2020dipolar,park2023dipole,lian2024valleypolarized,xiong2023correlated}.
Moreover, with optical control of exciton doping and gate control of carrier doping, recent experiments have realized Bose-Fermi mixtures for the exploration of their interplay in $\text{WS}_2/\text{WSe}_2$ heterobilayers~\cite{wang2022light,Wang_2023,gao2024excitonic}.

Magnetic interactions and magnetic orders are of paramount interest in the context of various TMDs moir\'e superlattices~\cite{tang2020simulation,wang2022light,Anderson_2023}. The discovery of intrinsic ferromagnetism in twisted bilayer MoTe$_2$~\cite{Anderson_2023} has heralded the experimental observations of integer and fractional quantum anomalous Hall effects in this exciting platform~\cite{cai2023signatures,park2023observation,zeng2023thermodynamic,xu2023observation}. Curie temperature sets a limit for the exploration of these exotic topological states of matter relying on broken time reversal symmetry. The existing observations in moir\'e platforms are reported in the order of $1-10$ K~\cite{serlin2020intrinsic, li2021quantum,cai2023signatures,park2023observation,zeng2023thermodynamic,xu2023observation}, likely limited by the density of the moir\'e carriers. 
Numerical studies suggest that the magnon gap decreases with lower carrier concentration, indicating the decreased stability of fractional quantum anomalous Hall phases at smaller filling fraction~\cite{PhysRevB.108.085117}.
This prompts the question of how to increase the stability of the magnetic order and the critical temperatures of the phenomena for easier experimental access, especially at low carrier filling fractions. 
Notably, experiments have discovered that,
in the dilute filling limit of a $\text{WS}_2/\text{WSe}_2$ moir\'e, light irradiation can help to establish ferromagnetic order~\cite{wang2022light}, where the role of optically injected mobile interlayer excitons has been identified as mediating a long-range RKKY type of magnetic interaction~\cite{xiao2023dynamic}.

\begin{figure}[pbth]
\begin{center}
\includegraphics[width=0.48\textwidth]{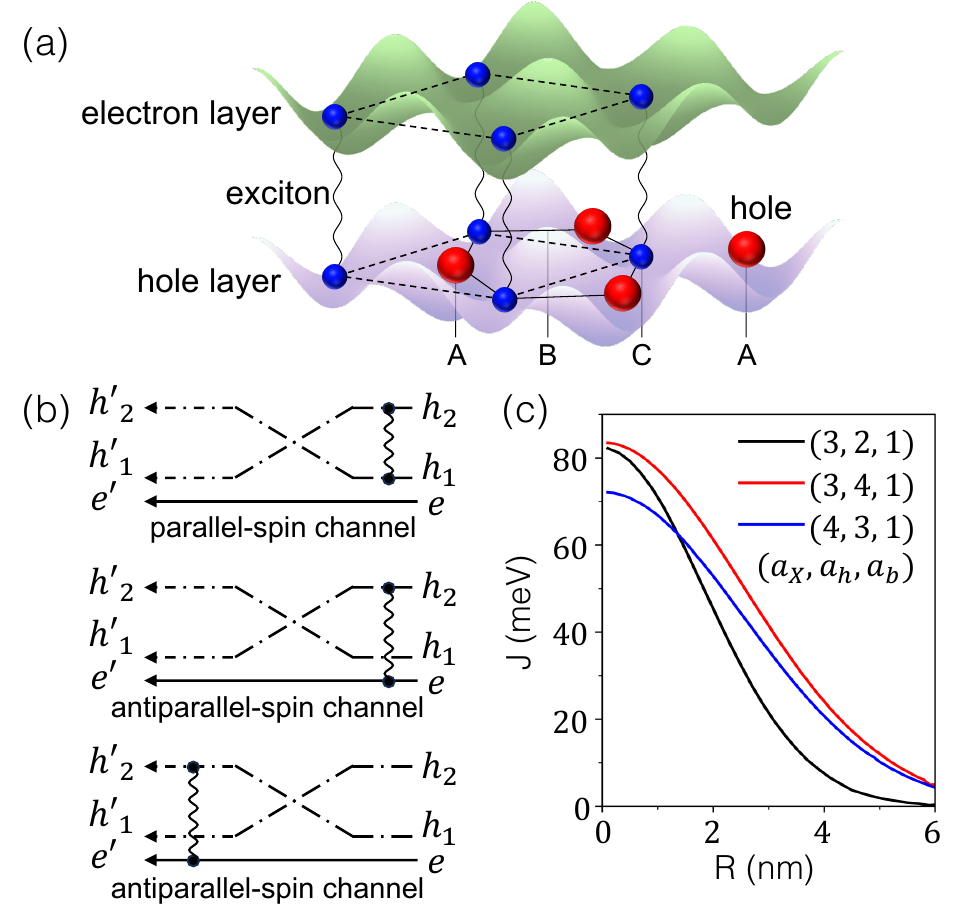}
\end{center}
\caption{(Color online)
(a) Schematic illustration of mixed Boson-Fermion (exciton-hole) mixtures in heterobilayer moir\'e superlattice,
where A, B, and C are the three high-symmetry points in the moir\'e unit cell.
The green and purple curved surfaces represent the moir\'e potential surfaces of the conduction and valence bands, respectively.
 The color of the balls corresponds respectively to their spin.
(b) Feynman diagrams of the three exciton-hole ($eh_1$-$h_2$) exchange channels.
The solid lines denote electrons, and the dotted lines denote holes.
The top channel favors parallel-spin alignment, while the bottom two channels favor antiparallel-spin alignment.
(c) Exciton-carrier exchange $J$ as a function of their distance, where $a_X, a_e$, and $a_b$ are the Bohr radius of the exciton-wave packet, hole-wave packet, and exciton, respectively (Appendix~\ref{wave-packet}). 
The positive $J$ value means the antiparallel-spin exchange dominates. $R$ is the distance between exciton and hole wave packet.
}
\label{fig:general}
\end{figure}

In this paper, we investigate the magnetic orders in the Bose-Fermi mixture of carriers and excitons both trapped in the moir\'e superlattices of TMDs heterostructures, with exciton density electrically tuned at fixed carrier filling. 
Wave-packet model analysis and self-consistent Hartree-Fock mean-field calculations
both find that the exchange interaction between a carrier and an interlayer exciton in the moir\'e traps leads to a sizable magnetic interaction, which, counterintuitively, are aligned in an antiparallel-spin configuration.
When carriers and excitons are trapped at different high symmetry points of the moir\'e supercell, their nearest neighbor interactions at a distance much shorter than the carrier-carrier separation establish a joint magnetic order in the Bose-Fermi mixture, where the carrier sublattice develops robust ferromagnetic order opposite to that in the exciton sublattice. 
The possibility of increasing the Curie temperature of moir\'e carriers is demonstrated through electrical tuning of the exciton density in the ground state.
In a trilayer moir\'e system with a p-n-p type band alignment, we show that the exciton-carrier interplay can establish a layered antiferromagnetism for holes confined in the two outer layers.
We further reveal a spontaneous nematic order in the Bose-Fermi mixture, arising from the interference between interlayer tunneling that creates electron-hole pair in p-wave channel and Coulomb interaction that favors s-wave channel instead.

\emph{\color{blue} Exciton-carrier exchange.}
Consider a type-II heterointerface at carrier filling $\nu=-1$, i.e. one hole per moir\'e unit cell (MUC).
Fig.~\ref{fig:general}(a) shows a moir\'e landscape in which the doped holes (red spheres) and interlayer excitons (pairs of blue spheres) are trapped at different high-symmetry points in the MUC, denoted as A and C sites.
As we will show by Hartree-Fock calculation, this is an energetically favorable configuration when holes and electrons have their energy minima at A and C sites respectively, like in the H-stacking WSe$_2$/WS$_2$ bilayer~\cite{Wang_2023,naik2022intralayer}, and when all A sites are occupied by a doped hole.
In such configuration, the exciton-carrier distance is much shorter compared with the carrier-carrier distance. The Coulomb exchange interaction between the carrier and exciton therefore becomes the dominant magnetic interaction that determines the ground state of the mixture.

This Bose-Fermi mixtures can be described by a Heisenberg model with exchange energy $J$ (Appendix~\ref{wave-packet}).
There are three exciton-hole exchange channels associated with this process, as illustrated in Fig.~\ref{fig:general}(b): 
one channel (top panel) contributes parallel-spin exchange, while the bottom two contribute antiparallel-spin exchanges.
The total exchange energy $J$ is determined by the competition among these three channels.

%
\begin{figure}[ptb]
\begin{center}
\includegraphics[width=0.48\textwidth]{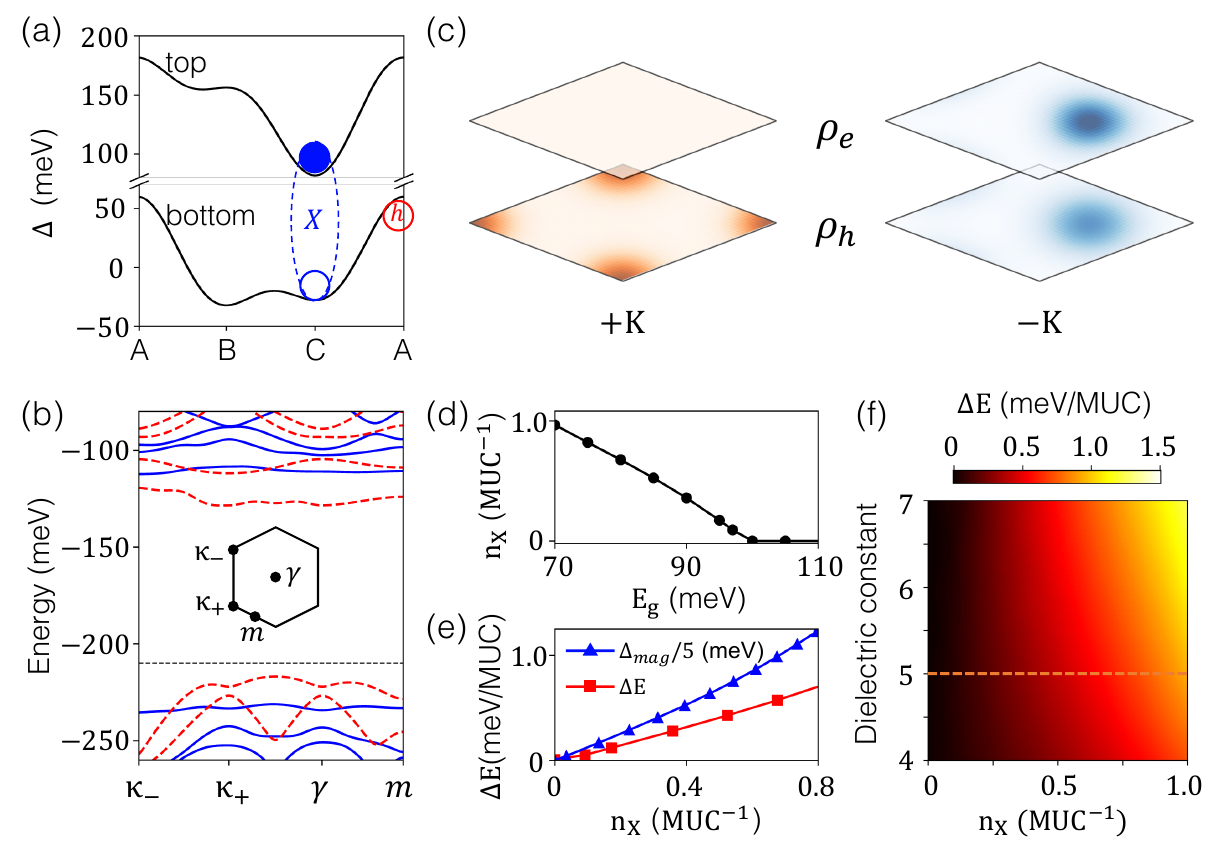}
\end{center}
\caption{(Color online)(a) Moir\'{e} potentials of top and bottom layers along the long diagonal of the moir\'{e} unit cell [Fig.~\ref{fig:general}(a)].
(b) Hartree-Fock band of heterobilayer moir\'{e} supperlattice at filling $\nu=-1$.
Dashed red (solid blue) lines are the bands associated with $K (-K)$ valley.
The black dashed line denote the Fermi level.
Parameters used are $m_t=0.45~m_e$, $m_b=-0.25~m_e$, $V_{t/b}=10~\text{meV}$, $\phi_{t/b}=45/-3^{\circ}$, $\epsilon_r=5$, $w=0$, $E_g=75~\text{meV}$.
(c) Layer resolved charge distribution of the ground state. All the doped holes (left panel) are spin up, locked with valley $+K$.
(d) Variation of exciton density as a function of energy gap.
(e) Variations of energy difference ($\Delta E \equiv E_{\mathrm{AFM}}-E_{\mathrm{FM}}$) and magnon gap ($\Delta_{mag}$) as functions of exciton density
(details in Appendix~\ref{hartree-fock}).
(f) Phase diagram of $\Delta E$ as a function of exciton density and relative dielectric constant. Dashed line indicates the cut shown in subfigure (d).
}
\label{fig:bilayer}
\end{figure}

We have calculated the total $J$ values for ballpark realistic parameters, including the exciton Bohr radius $a_b$, exciton wave-packet radius $a_X$, carrier wave-packet radius $a_h$, and exciton-carrier separation $R$.
This calculation assumes that the exciton center-of-mass wavefunction and carrier wavefunction are wave packets trapped by moir\'e potential (see details in Appendix~\ref{wave-packet}).
As shown in Fig.~\ref{fig:general}(c), $J$ is positive for a wide range of parameters, indicating an overall tendency toward antiparallel-spin alignment between excitons and carriers.

It has been established in a previous study~\cite{PhysRevLett.121.026402} that a carrier doped heterobilayer moir\'e superlattice can be described by Hubbard model,
 and with one hole MUC the system hosts weak in-plane anti-ferromagnetic (AFM) order among carriers.
We show here, supported by the analytic argument given above and more numerical evidence below,
 that a ferromagnetic (FM) order among carriers can be established with the aid of exciton.

\emph{\color{blue} Exciton mediated FM order in heterobilayer moir\'{e}. }
We first perform self-consistent Hartree-Fock (HF) calculations for heterobilayer moir\'e in the plane-wave basis to locate its ground state configurations.
The single-particle Hamiltonian (with its time-reversal copy) is
\begin{equation}
H_{\uparrow}=\begin{pmatrix}
\frac{(k-\kappa_{-})^2}{2m_t}+V_t(\bm{r})+E_g &T(\bm{r})\\
T^{\dagger}(\bm{r})&\frac{(k-\kappa_{+})^2}{2m_b}+V_b(\bm{r})
\end{pmatrix},
\end{equation}
where $V_{t/b}(\bm{r})=\sum_{i=1,3,5}2V_{t/b}cos(\bm{g_i}\cdot\bm{r}+\phi_{t/b})$ and $T(\bm{r})=w(1+\omega e^{i\bm{g_2}\cdot r}+\omega^2 e^{i\bm{g_3}\cdot r})$,
with $\bm{g_i}=b_{M}[\cos\frac{(i-1)\pi}{3}, \sin\frac{(i-1)\pi}{3}]$, $b_M$ is the moir\'e reciprocal lattice constant, $\omega=e^{\frac{i2n\pi}{3}}$, in which $n$ is determined by angular momentum difference between orbitals involved.
The moir\'e potential employed in the calculations for the heterobilayer system is ballpark realistic with $|V_{b/t}|=10~\text{meV}$ and their profile is 
a typical one capturing the feature that holes and electrons have their energy minima at different high symmetry points in an MUC,
as shown in Fig.~\ref{fig:bilayer}(a).
We take moir\'e period 10~nm,
and TMD lattice constant $a_0 = 0.347~\mathrm{nm}$  throughout this paper. 

We focus on the scenario with filling factor $\nu=-1$, i.e., one hole per MUC. 
In the calculation, the type-II band gap $E_g$ is taken to be slightly smaller compared to the interlayer exciton binding energy, so that Coulomb interaction leads to the spontaneous formation of excitons.
The real-space charge density of ground state is shown in Fig.~\ref{fig:bilayer}(c) and the corresponding band structure in Fig.~\ref{fig:bilayer}(b).
The doped hole is trapped in the bottom layer (valence band) around A point, while the interlayer exciton is trapped around C point, corresponding to the scenario of Fig.~\ref{fig:general}(a).
In addition, they have opposite spin/valley quantum number.
Thus, we observe a scenario in which the spontaneously formed exciton and doped carriers exhibit antiparallel-spin alignment.

We proceed to investigate the tunability and robustness of the observed magnetic ground state.
In experiments, the exciton density may be tuned by the light intensity in the optically injected scenario~\cite{wang2022light}, and by electrostatic gating in the
co-doping scenario~\cite{wang2019evidence}. In the HF calculation, the density of the exciton is practically controlled by the band gap $E_g$ that is tunable by the out-of-plane electric field.
As $E_g$ is decreased, the exciton density grows.
The phase indicator $\Delta E$  is defined as the energy difference between two magnetic configurations (Appendix~\ref{hartree-fock}):
 (1) FM state with all carriers being spin up locked with valley $+K$,
 (2) AFM state with half of the carriers spin up, while the other half spin down, i.e., $\Delta E \equiv E_{AFM}-E_{FM}$.
 We expand the system to $1\times 2$ MUCs when performing the HF calculation, and the details are laid out in Appendix~\ref{hartree-fock}.

As shown in Fig.~\ref{fig:bilayer}(d), when $E_g$ is tuned lower than a critical point, exciton spontaneously condenses (black line), the system transforms to an exciton insulating state, and as shown in Fig.~\ref{fig:bilayer}(e), $\Delta E$ (red line) deviates from zero concurrently with increasing exciton density.
This indicates that the emergence of excitons facilitates the establishment of a magnetic order, specifically ferromagnetism, within the carrier sublattice.
The full phase diagram is shown in Fig.~\ref{fig:bilayer}(f) with various choices of exciton density $n_X$ and dielectric constant $\epsilon_r$. 
A positive phase indicator $\Delta E$ always occurs together with a non-zero exciton density $n_X$, and their magnitude is positively correlated, showing a robust exciton-mediated magnetic order in the heterobilayer moir\'e system.

We further carry out exact diagonalization on top of the self-consistent ground state, and calculate the magnon gap $\Delta_{mag}$ associated with flipping one hole's spin/valley (Appendix~\ref{hartree-fock}).
A non-zero magnon gap occurs simultaneously with the non-zero exciton density, as shown by blue-dot line Fig.~\ref{fig:bilayer}(e).
Both the $\Delta E$ and $\Delta_{mag}$ are positively correlated with $n_X$, so the injected exciton can help to stabilize the FM order among the doped carriers.
This may be exploited for exploring correlation phenomena that rely on the ferromagnetic order of moir\'e carriers.

\begin{figure}[pt]
\begin{center}
\includegraphics[width=0.48\textwidth]{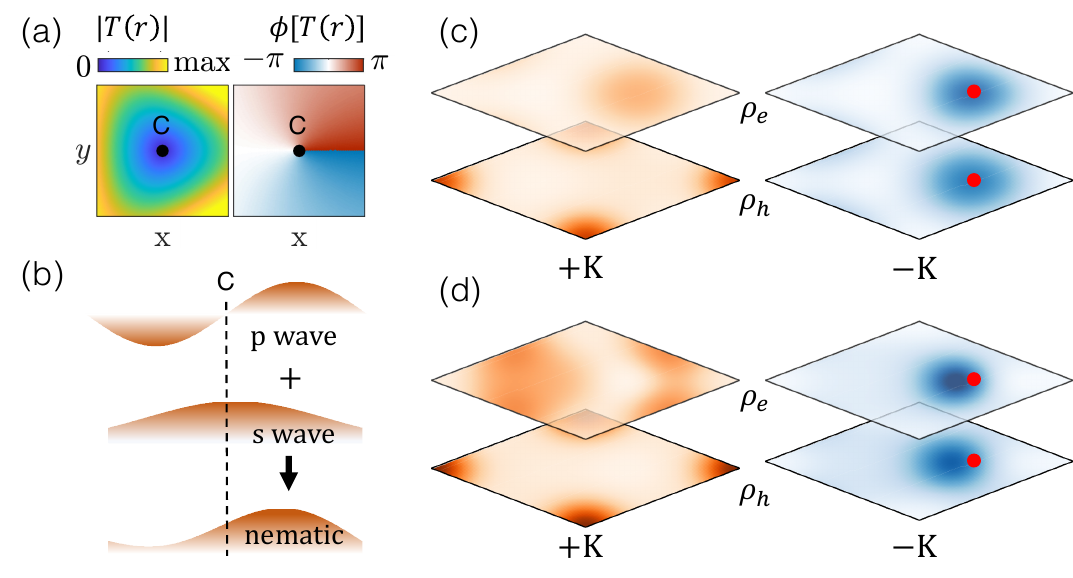}
\end{center}
\caption{(Color online)
(a) Schematic illustration of the p-wave tunneling in real space. The left (right) panel is the tunneling magnitude (phase).
(b) Schematic illustration of nematic order induced by s-p interference, where the dashed line denotes the rotation axis (high symmetry point).
(c, d) Layer resolved charge distribution of $+K$ (upper panel) and $-K$ (lower panel) valley.
The red points denote the high-symmetry point C.
In (c), the interlayer tunneling $T(\bm{r})$ has an s-wave form in the vicinity of C point.
While in (d), a different choice of $\omega$ leads to an interlayer tunneling $T(\bm{r})$ that has p-wave form in the vicinity of C point. As evident from the density profile, the $C_3$ symmetry is broken, nematic order emerges.
The parameters used are the same as in Fig.~\ref{fig:bilayer} except $w=3~\text{meV}$.
}
\label{fig:nematic}
\end{figure}
\emph{\color{blue} Nematic order in heterobilayer moir\'{e}.}
We then focus on the interplay between the Coulomb interaction and interlayer tunneling, both of which can lead to the formation of interlayer electron-hole pair.
Coulomb interaction leads to exciton condensation in the s-wave channel, with a spontaneously determined global phase.
On the other hand, the interlayer tunneling in the vicinity of the high-symmetry stackings can have a p-wave symmetry~\cite{zhu2019gate}. The $C_3$ rotational symmetry dictates that interlayer tunneling coefficient $T(\bm{r})$ must have p-wave symmetry in the vicinity of two out of the three high symmetry points of the MUC [c.f. Fig.~\ref{fig:nematic}(a)]~\cite{tong2017topological}, whereas at the remaining site, $T(\bm{r})$ has the s-wave form. If exciton is formed at p-wave sites, the p-wave interlayer tunneling leads to electron-hole pair formation in the p-wave channel with a fixed global phase, and its interference with the Coulomb induced pairing in s-wave channel can shift the exciton wavefunction away from the high symmetry point, thereby break the $C_3$ rotation symmetry and introducing into nematic order~\cite{zhu2019gate}. This is schematically illustrated in Fig.~\ref{fig:nematic}(b).

Motivated by these considerations, we carry out HF calculation in the same heterobilayer system as in last section, with the interlayer tunneling $T(\bm{r})$ turned on.
The resulting real-space density profile is shown in Fig.~\ref{fig:nematic}(c, d).
In Fig.~\ref{fig:nematic}(c), 
$T(\bm{r})$ takes a form that has s-wave form in the vicinity of C site.
As Coulomb interaction and interlayer tunneling both favor electron-hole pairing in the s-wave channel, the $C_3$ rotational symmetry is retained in the ground state.
For the calculation in Fig.~\ref{fig:nematic} (d), a different choice of parameter $\omega$ leads to a $T(\bm{r})$ that has p-wave form in the vicinity of C site.
Compared with the density profile presented in Fig.~\ref{fig:bilayer}(b) (where interlayer tunneling is switched off) and Fig.~\ref{fig:nematic}(c), apparent deviation of the center of exciton from the high-symmetry C point is observed.
We also remark that in the pure-Coulomb case presented in Fig.~\ref{fig:bilayer}(b) (without interlayer tunneling), the spontaneously formed exciton is fully polarized in one valley.
However, as we turn on the tunneling, the exciton is not fully polarized as interlayer tunneling occurs in both valleys, as shown  in Fig.~\ref{fig:nematic} (c, d). Nevertheless, the ground state still prefers partial polarization where most excitons are in the opposite valley to the doped holes.

\begin{figure}[pt]
\begin{center}
\includegraphics[width=0.48\textwidth]{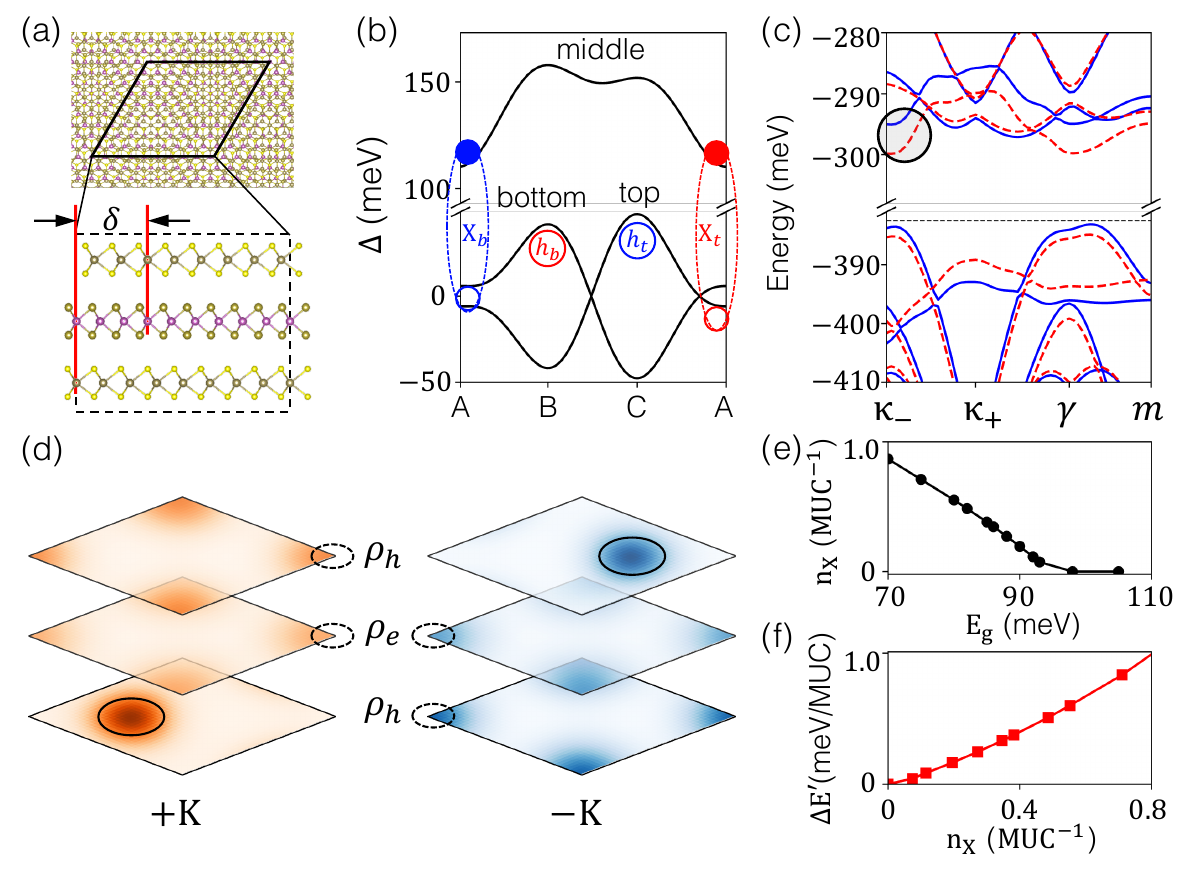}
\end{center}
\caption{(Color online) (a) Schematic illustration of sandwich-like trilayer moir\'e superlattice.
The upper (lower) panel is the top (side) view, $\delta$ is the relative displacement between the top and bottom layers.
(b) Moir\'e potentials of the sandwich-like moir\'e superlattice along the long diagonal of the moir\'e unit cell.
(c) Hartree-Fock quasiparticle bands of the two-hole doped trilayer moir\'{e} superlattice, the black circle marks the two hole bands.
Parameters used are $m_{t/b}=-0.25~m_e$, $m_{m}=0.33~m_e$, $V_{t/b}=8.6~\text{meV}$, $V_m=5.0~\text{meV}$, $\phi_{t/m/b}=-96/173/84^{\circ}$~\cite{tong2020Interferences}, $E_g=75~\text{meV}$, $\epsilon_r=5$.
(d) Layer resolved charge distribution of $+K$ (left) and $-K$ (right) valley, in which the solid (dashed) circles denote the carrier holes (excitons).
(e-f) Variation of (e) exciton density and (f) energy difference between the layered FM and AFM states as a function of energy gap and exciton density, respectively.
}
\label{fig:trilayer}
\end{figure}
\emph{\color{blue} Layered AFM order in trilayer moir\'{e}.}
An immediate extension of the results established above is to multi-layer system.
Consider a sandwich-like trilayer moir\'e heterostructure, as shown in Fig.~\ref{fig:trilayer}(a), with the top and bottom layer (middle layer) hosting valence (conduction) band.
If doped with holes, these carriers will be located in the top and bottom layers, and there is no magnetic interaction between them  owing to the lack of spatial overlap.

The moir\'e potentials here depend on the relative interlayer translation [Fig.~\ref{fig:trilayer}(a)] and orientation between layers~\cite{tong2020Interferences}.
A different configuration may yield different results, but the underlying physics remains the same.
For simplicity, we focus on the scenario where the interlayer excitons from two interfaces are trapped in the same horizontal location [Fig.~\ref{fig:trilayer}(b)].
In this configuration, two exciton holes (hollow circles) are well separated into the two outer layers,
 while the two exciton electrons (solid circles) exhibit a strong overlap in the middle layer.
Because of the strong overlap between the two electrons in the middle layer, the Pauli exclusion forces these two excitons to be spin-antiparallel aligned~\cite{huang2024nonbosonic}.
With the exciton-carrier antiparallel-spin exchange interactions in the top and bottom layers, even when separated by a large distance, a layered AFM order can be established between the carriers trapped in the top and bottom layers.

This is verified by HF calculation in trilayer moir\'e with ballpark realistic parameters at filling factor $\nu=-2$.
As shown by the HF quasiparticle bands in Fig.~\ref{fig:trilayer}(c),
 the two doped holes (marked by the black circle) are AFM aligned in valley/spin.
The layer-resolved charge densities are shown in Fig.~\ref{fig:trilayer}(d).
Similar to the bilayer scenario, the interlayer excitons (dashed circles) in the upper (lower) moir\'e interface are antiparallel-spin aligned with the holes in the top (bottom) layer,
 while the two carrier holes (solid circles) are positioned individually in the top and bottom layers.
Thus, a layered AFM order is established between the carriers in the two outer layers despite the lack of wave function overlap between the holes in two distinct layers.

To verify that the observed layered magnetic order is mediated by exciton, as shown in Fig.~\ref{fig:trilayer}(e), we tune the exciton density $n_X$ by decreasing the band gap $E_g$ across the critical point.
The phase indicator is defined as the energy difference between the layer-AFM state and layer-FM state, $\Delta E^\prime = E_{LFM} - E_{LAFM}$.
As presented in Fig.~\ref{fig:trilayer}(f), $\Delta E^\prime$ vanishes simultaneously with $n_X$, i.e., without exciton, there is no magnetic order in the system.
While we have chosen a special configuration of hetero-trilayer moir\'e to present, this mechanism can establish layered magnetic order in more general configurations.
The magnetic order between two outer layers will be determined not only by the exchange between carriers and excitons, but also by the interplay between excitons in the two moir\'e interfaces.
In a general incommensurate trilayer system, the structure of this exchange requires more detailed study, which we leave for the future.

\emph{\color{blue} Conclusion and discussion.} 
In this article, we have shown that exciton can help to establish and stabilize an effective ferromagnetic exchange between carriers widely separated in space.
As such, externally injected exciton can be used to introduce long-range magnetic order in moir\'e superlattice.
More practically, they may help to raise the Curie temperature in moir\'e superlattice, which is crucial, for example, for exploring the correlation phenomena that rely on magnetic order, such as the FQAH state.
We limited our study to only a few set of parameters in this article. However, it is evident that the mechanism reported is universal in the large moir\'e family.
While the calculations are performed for a ground state exciton-carrier mixture by tuning exciton density with a band gap parameter $E_g$ much smaller than that in realistic compounds, the revealed mechanism of forming magnetic order through exciton-carrier exchange generally applies to systems with excitons injected either optically~\cite{wang2022light} or electrically~\cite{wang2019evidence}.
We tune the gap parameters in Hartree-Fock calculation such that the Coulomb excitons have a density close to that of excitons injected by light in experiments [$O(0.1/\mathrm{MUC})$].

The study presented in this article can be generalized immediately in a few aspects.
For example, in trilayer moir\'e, we have chosen two outer layers to be aligned for simplicity, so there is only one moir\'e scale present in the problem.
However, in realistic trilayer system, the two interfaces can have distinct moir\'e periodicity, leading to a bichromatic landscape for the Bose-Fermi mixture~\cite{tong2020Interferences}.
The consequence of the antiparallel-spin exchange between interlayer excitons and carriers in such complex landscapes can be an interesting subject of study.

\emph{\color{blue} Acknowledgement.} T. Tan acknowledges useful conversation with T. Devakul, J. Dong, A. P. Reddy, Yves Kwan. This work is supported by the National Key R\&D Program of China (2020YFA0309600), Research Grant Council of Hong Kong SAR (AoE/P-701/20, HKU SRFS2122-7S05), and New Cornerstone Science Foundation. T. Tan is supported by Stanford Graduate Fellowship. 

\appendix
\section{Hartree Fock Calculation\label{hartree-fock}}
We performed Hartree-Fock calculation in plane-wave basis in TMD heterbilayer/trilayer. We detailed the operational perspective of the calculation in this appendix.

\subsection{Single Particle Hamiltonian}
The single-particle Hamiltonian of moir\'e TMD heterobilayer with Type-II band alignment for spin up (locked with $+K$ valley) is
\begin{equation}
H_{\uparrow} =\begin{pmatrix}
\frac{(k-\kappa_{-})^2}{2m_t}+V_t(\bm{r})+E_g &T(\bm{r})\\
T^{\dagger}(\bm{r})&\frac{(k-\kappa_{+})^2}{2m_b}+V_b(\bm{r})
\end{pmatrix},
\end{equation}
Its time reversal partner (spin down locked with $-K$ valley) is
\begin{equation}
H_{\downarrow} =\begin{pmatrix}
\frac{(k+\kappa_{-})^2}{2m_t}+V_t(\bm{r})+E_g &T^{\dagger}(\bm{r})\\
T(\bm{r})&\frac{(k+\kappa_{+})^2}{2m_b}+V_b(\bm{r})
\end{pmatrix},
\end{equation}
 where $V_{t/b}(\bm{r})=\sum_{i=1,3,5}2V_{t/b}cos(\bm{g_i}\cdot\bm{r}+\phi_{t/b})$, $T(\bm{r})=w(1+\omega e^{i\bm{g_2}\cdot r}+\omega^2 e^{i\bm{g_3}\cdot r})$,
with $\bm{g_i}=b_{M}[\cos\frac{(i-1)\pi}{3}, \sin\frac{(i-1)\pi}{3}]$, $b_M$ is the moir\'e reciprocal lattice constant, $\omega=e^{\frac{i2n\pi}{3}}$, where $n$ is determined by angular momentum difference between orbitals involved.

For trilayer system, we take the single particle Hamiltonian to describe a sandwitch-like structure, with outer two layers contributing valence band, and the middle layer contributing the conduction band. The single-particle Hamiltonians in the $+K$ and $-K$ valleys are
\begin{widetext}
\begin{align}
\begin{split}
H_{\uparrow}^T &=\begin{pmatrix}
\frac{(k-\kappa_{-})^2}{2m_t}+V_t(\bm{r}) &T(\bm{r})&0\\
T^{\dagger}(\bm{r})&\frac{(k-\kappa_{+})^2}{2m_m}+V_m(\bm{r})&T^{\dagger}(\bm{r})\\
0&T(\bm{r})&\frac{(k-\kappa_{-})^2}{2m_b}+V_b(\bm{r})
\end{pmatrix}
-E_g\begin{pmatrix}
1&0&0\\
0&0&0\\
0&0&1
\end{pmatrix},
\end{split}
\\[3ex]
\begin{split}
H_{\downarrow}^T &= \begin{pmatrix}
\frac{(k+\kappa_{-})^2}{2m_t}+V_t(\bm{r}) &T^{\dagger}(\bm{r})&0\\
T(\bm{r})&\frac{(k+\kappa_{+})^2}{2m_m}+V_m(\bm{r})&T(\bm{r})\\
0&T^{\dagger}(\bm{r})&\frac{(k+\kappa_{-})^2}{2m_b}+V_b(\bm{r})
\end{pmatrix}
-E_g\begin{pmatrix}
1&0&0\\
0&0&0\\
0&0&1
\end{pmatrix}.
\end{split}
\end{align}
\end{widetext}
For Figs.~\ref{fig:bilayer} and \ref{fig:trilayer}, we set $w=0$ to focus on the s-wave channel exciton. When studying the nematic effect [Fig.~\ref{fig:nematic}(a)], we set $w=3$~meV.

\subsection{Hartree-Fock Calculation}
Let $c^{\dagger}_{\bm{k},\bm{b},\tau,l}$ be the electron creation operator in plane wave state with momentum $\bm{k}+\bm{b}$, measured relative to $\gamma$ point, $\bm{b}$ is a moir\'e reciprocal lattice vector with basis $\bm{g}_i$, $\tau$ labels valley/spin (locked), $l$ labels layer.

The many-body Hamiltonian is written as
\begin{widetext}
\begin{equation}
\begin{split}
\hat{H}&=\sum_{\substack{\bm{k}\in mBZ\\\bm{b}_1,\bm{b}_2\\l_1,l_2,\tau}}T_{\bm{b}_1 l_1,\bm{b_2}l_2}(\bm{k},\tau)c^{\dagger}_{\bm{k},\bm{b_1},l_1,\tau}c_{\bm{k},\bm{b_2},l_2,\tau}\\
 &\quad +\frac{1}{2A}\sum_{\substack{\bm{k_1}-\bm{k_4}\in mBZ\\\bm{b}_1-\bm{b}_4\\l_1 l_2,\tau_1,\tau_2}} V_{l_1l_2}(\bm{k}_1+\bm{b}_1-\bm{k}_3-\bm{b}_3) \\
&\quad \quad \times c^{\dagger}_{\bm{k_1},\bm{b}_1,l_1,\tau_1}c^{\dagger}_{\bm{k_2},\bm{b}_2,l_2,\tau_2}c_{\bm{k_4},\bm{b}_4,l_2,\tau_2}c_{\bm{k_3},\bm{b}_3,l_1,\tau_1}
\delta(\bm{k_1}+\bm{b}_1+\bm{k}_2+\bm{b}_2-\bm{k}_3-\bm{b}_3-\bm{k}_4-\bm{b}_4),
\end{split}
\end{equation}
\end{widetext}
where $T_{\bm{b}_1 l_1,\bm{b_2}l_2}(\bm{k},\tau)$ is the matrix element of the single particle Hamiltonian $H_{\tau}$ (defined above) in the plane wave basis, A is sample area. $V(\bm{q})_{l_1,l_2}$  is the Fourier-transform of the Coulomb interaction between layer $l_1$ and $l_2$, the delta function ensures momentum conservation. We take dual-gate screened Coulomb interaction and neglect inter-valley scattering.
\begin{widetext}
\begin{equation}
V(\bm{q})_{l_1,l_2}=\frac{e^2}{2\epsilon_0\epsilon_r|\bm{q}|}\left[\exp(-d|\bm{q}|)-\frac{2\exp(-2D|\bm{q}|)}{1+\exp(-2D|\bm{q}|)}\right],
\end{equation}
\end{widetext}
$d$ is vertical distance between layer $l_1$ and $l_2$. We take inter-layer distance to be $0.7$nm. $D$ is the gate-to-sample distance, we take $D=10$nm.

Hartree-Fock approximation consists of substituting the four-fermion operator by its pair-wise contraction
\begin{widetext}
\begin{align}
\begin{split}
&c^{\dagger}_{\bm{k_1},\bm{b}_1,l_1,\tau_1} c^{\dagger}_{\bm{k_2},\bm{b}_2,l_2,\tau_2}c_{\bm{k_4},\bm{b}_4,l_2,\tau_2}c_{\bm{k_3},\bm{b}_3,l_1,\tau_1}\\
&\to
 c^{\dagger}_{\bm{k_1},\bm{b}_1,l_1,\tau_1}c_{\bm{k_3},\bm{b}_3,l_1,\tau_1}\braket{c^{\dagger}_{\bm{k_2},\bm{b}_2,l_2,\tau_2}c_{\bm{k_4},\bm{b}_4,l_2,\tau_2}}
 + \braket{c^{\dagger}_{\bm{k_1},\bm{b}_1,l_1,\tau_1}c_{\bm{k_3},\bm{b}_3,l_1,\tau_1}}c^{\dagger}_{\bm{k_2},\bm{b}_2,l_2,\tau_2}c_{\bm{k_4},\bm{b}_4,l_2,\tau_2}\\
 & -\braket{c^{\dagger}_{\bm{k_1},\bm{b}_1,l_1,\tau_1}c_{\bm{k_4},\bm{b}_4,l_2,\tau_2}}c^{\dagger}_{\bm{k_2},\bm{b}_2,l_2,\tau_2}c_{\bm{k_3},\bm{b}_3,l_1,\tau_1}
-c^{\dagger}_{\bm{k_1},\bm{b}_1,l_1,\tau_1}c_{\bm{k_4},\bm{b}_4,l_2,\tau_2}\braket{c^{\dagger}_{\bm{k_2},\bm{b}_2,l_2,\tau_2}c_{\bm{k_3},\bm{b}_3,l_1,\tau_1}}.
\end{split}
\end{align}
\end{widetext}
The $\langle \rangle$ is taken with respect to the Slater determinant ground state constructed from the Hartree-Fock orbitals. We allow symmetry breaking at the moir\'e scale, and do not allow inter-valley coherence.
\begin{equation}
\braket{c^{\dagger}_{\bm{k_1},\bm{b}_1,l_1,\tau_1}c_{\bm{k_2},\bm{b}_2,l_2,\tau_2}}\propto \delta_{\bm{k}_1,\bm{k}_2}\delta_{\tau_1,\tau_2},
\end{equation}
After this substitution, the Hartree-Fock Hamiltonian is
\begin{widetext}
\begin{align}
\begin{split}
\hat{H}_{H}(\mathcal{P})
&=\sum_{\substack{\bm{k}_1,\bm{k}_2\\l_1,l_2,\tau_1,\tau_2\\\bm{b}_1-\bm{b}_4}}c^{\dagger}_{\bm{k}_1,\bm{b}_1,l_1,\tau_1}c_{\bm{k}_1,\bm{b}_3,l_1,\tau_1}\mathcal{P}(\bm{k}_2,\tau_2)_{\bm{b}_4 l_2,\bm{b}_2 l_2}
 \frac{V_{l_1,l_2}(\bm{b}_1-\bm{b}_3)}{A}\delta(\bm{b}_1+\bm{b}_2-\bm{b}_3-\bm{b}_4)\\
&\equiv \sum_{\substack{\bm{k}_1\\l_1,\tau_1\\\bm{b}_1 \bm{b}_3}} c^{\dagger}_{\bm{k}_1,\bm{b}_1,l_1,\tau_1}[H_{H}(\bm{k_1},\tau_1)_{\bm{b}_1 l_1,\bm{b}_3 l_1}]c_{\bm{k}_1,\bm{b}_3,l_1,\tau_1},
\end{split}
\\[2ex]
\begin{split}
\hat{H}_{F}(\mathcal{P})&=-\sum_{\substack{\bm{k}_1,\bm{k}_2\\l_1,l_2,\tau_1\\\bm{b}_1-\bm{b}_4}}c^{\dagger}_{\bm{k_1},\bm{b}_1,l_1,\tau_1}c_{\bm{k_1},\bm{b}_4,l_2,\tau_1}\mathcal{P}(\bm{k}_2,\tau_1)_{\bm{b}_3 l_1,\bm{b}_2l_2}
 \frac{V_{l_1,l_2}(\bm{k_1}+\bm{b_1}-\bm{{k}_2-\bm{b}_3)}}{A} \delta(\bm{b}_1+\bm{b}_2-\bm{b}_3-\bm{b}_4)\\
&\equiv 
\sum_{\substack{\bm{k}_1\\l_1,l_2,\tau_1\\\bm{b}_1,\bm{b}_4}} c^{\dagger}_{\bm{k_1},\bm{b}_1,l_1,\tau_1}[H_{F}(\bm{k}_1,\tau_1)_{\bm{b}_1 l_1,\bm{b}_4 l_2}]c_{\bm{k_1},\bm{b}_4,l_2,\tau_1},
\end{split}
\\[2ex]
\begin{split}
\hat{H}_{HF}&=\hat{H}_H(\mathcal{P}-\mathcal{P}_{CNP})+\hat{H}_F(\mathcal{P}-\mathcal{P}_{CNP})
 +\sum_{\substack{\bm{k}\in mBZ\\\bm{b}_1,\bm{b}_2\\l_1,l_2,\tau}}T_{\bm{b}_1 l_1,\bm{b_2}l_2}(\bm{k},\tau)c^{\dagger}_{\bm{k},\bm{b_1},l_1,\tau}c_{\bm{k},\bm{b_2},l_2,\tau},
\end{split}
\end{align}
\end{widetext}
where we have defined the projector $\mathcal{P}$ as
\begin{equation}
\mathcal{P}(\bm{k},\tau)_{\bm{b}_1 l_1,\bm{b}_2 l_2}\equiv \braket{c^{\dagger}_{\bm{k},\bm{b}_2,l_2,\tau}c_{\bm{k},\bm{b}_1,l_1,\tau}}.
\end{equation}
$\mathcal{P}_{CNP}$ refers to projector evaluated with respect to the rotated isolated TMD hetero bilayer/trilayer filled to charge neutrality point(CNP), i.e., no interlayer tunneling $T(\bm{r})$ and  moir\'e potential $V(\bm{r})$, only the kinetic term). This is to subtract off the charge background at charge neutrality \cite{PhysRevLett.124.097601}. 

The Hartree-Fock orbitals are eigenstates of $\hat{H}_{HF}$
\begin{equation}
d^{\dagger}_{\bm{k},n,\tau}=\sum_{\bm{g},l}z^{n\bm{k}\tau}_{\bm{g},\bm{l}}c^{\dagger}_{\bm{k},\bm{g},l,\tau},
\end{equation}
where $z$ is obtained by diagonalizing $\hat{H}_{HF}$ in the $c^{\dagger}, c$ basis, $n$ labels the Hartree-Fock bands. The Slater determinant ground state is
\begin{equation}
\ket{\text{Slater}}=\prod_{(n,\bm{k},\tau)\in \text{filled}}d^{\dagger}_{\bm{k},n,\tau}\ket{\Omega}.
\end{equation}
Then the projector is
\begin{equation}
\mathcal{P}(\bm{k},\tau)_{\bm{b}_1l_1,\bm{b}_2l_2}=\sum_{n\in \text{filled}}z^{n\bm{k}\tau}_{\bm{b}_1l_1}z^{*n\bm{k}\tau}_{\bm{b}_2l_2}.
\end{equation}

Thus, $H_{HF}$ and $\mathcal{P}$ depends on each other: We need $\mathcal{P}$ to construct the $H_{HF}$. We need  eigenvectors of $H_{HF}$ to construct $\mathcal{P}$. They are solved iteratively until convergence. The self-consistency condition is that the projector constructed from two iterative iteration are the same. We take the $\bm{k}$ points to be
\begin{equation}
\bm{k}=\frac{n_1}{3}\bm{g}_1+\frac{n_2}{3}\bm{g}_2,
\end{equation}
where $n_i=0,1,2$. We keep all the $|\bm{b}|\leq 3|\bm{g}|$ in our calculation. Finally, notice that the total energy of the state is
\begin{widetext}
\begin{equation}
\begin{split}
E &=\sum_{\bm{k},\tau}Tr\left\{\left[ T(\bm{k,\tau})+\frac{H_{H}(\bm{k},\tau)+H_{F}(\bm{k},\tau)}{2}\right]
 \times \left[ \mathcal{P}(\bm{k},\tau)-\mathcal{P}(\bm{k},\tau)_{CNP}\right]\right\}.
\end{split}
\end{equation}
\end{widetext}

A initial guess for the projector is feed into the program for the first iteration. We use a variety of initial projector, including random seed and ansatz motivated by physical consideration, to locate the lowest-energy state.

\subsection{Expanding the moir\'e unit cell}
\begin{figure}[ptb]
\begin{center}
\includegraphics[width=0.48\textwidth]{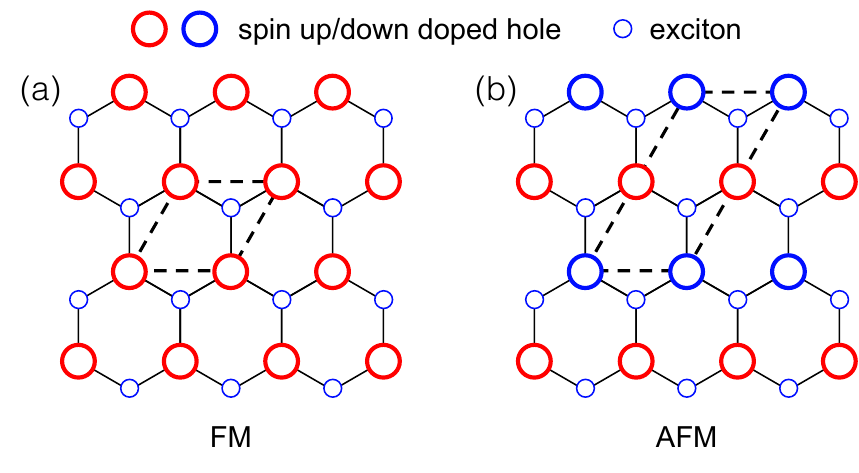}
\end{center}
\caption{(Color online)
Schematic illustration of the two magnetic states used in computing the phase indicator $\Delta E$.
The large (small) circles represent the doped holes (exciton), with the color indicating the spin species: red for spin up and blue for spin down.
(a) FM state: All doped holes (red circles) are in the spin-up state.
(b) AFM state: Half of the doped holes (red circles) are in the spin-up state, while the other half (blue circles) are in the spin-down state.
The parallelograms (dashed lines) in both figures represent the unit cells of the respective systems.
}
\label{SPfig_schematic}
\end{figure}

We mentioned in the main text that our Hartree-Fock calculation for the heterobilayer system is performed by expanding the moir\'e unit cell. Let $\bm{a}_1$ and $\bm{a}_2$ be the moir\'e vectors spanning one MUC, and $\bm{g}_1$ and $\bm{g}_2$ be the corresponding moir\'e reciprocal lattice vectors satisfying $\bm{a}_i \cdot \bm{g}_j=\delta_{ij}2\pi$.
We define the super moir\'e unit cell (SMUC) spanned by $\bm{a}_1$ and $2\bm{a}_2$. Each SMUC contains two MUC, which we will refer to MUC-1 and MUC-2. The corresponding super moir\'e reciproal lattice is spanned by $\bm{g}_1$ and $\frac{\bm{g}_2}{2}$.

The motivation for doing this expansion is that we want to investigate what magnetic order is preferred by the carriers, and whether the order is established by the exciton. Our claim in the main text is that the exciton helps to stabilize the ferromagnetic order among carries. Performing Hartree-Fock calculation at $\nu=-1$ without expansion, we cannot access the energy of the state where carriers of neighboring MUC have different spin/valley quantum number.

We performed calculation with two holes per SMUC, and located several states that are local energy minima. The ground state corresponds to the case where the transnational symmetry at the MUC scale is unbroken, with all carries in one valley, and excitions in the other valley. This is the case whose charge density is plotted in Fig.~\ref{fig:bilayer}(b). We refer to this state as FM state all carriers (holes) are in FM order, as shown in Fig.~\ref{SPfig_schematic}(a).
In contrast, the AFM state we referred to in Figs.~\ref{fig:bilayer}(e) and \ref{fig:bilayer}(f) is the following: half of the carriers are in valley $K$, and the other half in $-K$, as shown in Fig.~\ref{SPfig_schematic}(b). So, the energy difference between these two states, as shown by EQ~(\ref{eq:deltae}), can be used as phase indicator,
\begin{equation}
\Delta E \equiv E_{AFM}-E_{FM} \label{eq:deltae}
\end{equation}

The vanishing of the phase indicator $\Delta E$ at large $E_g$ (vanishing $n_X$) proves that without exciton, the large spatial separation between carriers quenches the magnetic order, and it is the onset of excitons that helps to mediate the long-range magnetic order among carriers.
We stress that we used a variety of initial projectors to locate the global energy minimum, which is the FM state. 

\subsection{Magnon Gap}
We presented in the main text the magnon gap, which is the energy of flipping one electron's spin/valley. There are subtleties associated with how to use the wave function/energy obtained from Hartree-Fock calculation to do exact diagonalization without double counting the interaction. We detail that in this subsection.

The Hartree-Fock orbitals are denoted $d^{\dagger}_{i}$ [to avoid cluttered notation and for generality, we use $i$ to represent the triplet $(\bm{k},n,\tau)$ used in previous sections]. The Hartree-Fock calculation is performed at certain filling (represented by $occ$, referring to the occupied orbitals). The following relation holds for the Hartree-Fock orbitals~\cite{Giuliani_Vignale_2005}:
\begin{widetext}
\begin{equation}
T_{ij}+\sum_{n\in occ}(V_{injn}-V_{innj})-\sum_{\gamma \in CNP}(V_{i\gamma j\gamma}-V_{i\gamma \gamma j})=\epsilon_i \delta_{ij},
\end{equation}
\end{widetext}
where $T_{ij}$ is the matrix element of the single-particle part of the Hamiltonian, and $V_{ijkl}$ is the matrix element of the interaction, evaluated in the Hartree-Fock orbital basis. $\epsilon_i$ is the eigenvalue associated with the orbital $d^{\dagger}_{i}$.
$\gamma$ is associated with the subtraction scheme discussed earlier. It refers to the eigenstates of the isolated rotated heterobilayer/trilayer structure. These $\gamma$ orbitals are filled to the charge neutrality point.

The many-body Hamiltonian, written in the Hartree-Fock orbital basis, is
\begin{widetext}
\begin{equation}
\begin{split}
\hat{H}_{d} = &\sum_{ij}\left[T_{ij}-\sum_{\gamma \in CNP}(V_{i\gamma j\gamma}-V_{i\gamma \gamma j})\right]d^{\dagger}_id_j
 +\frac{1}{2}\sum_{ijkl }V_{ijkl}d^{\dagger}_{i}d^{\dagger}_jd_{l}d_k.
\end{split}
\end{equation}
\end{widetext}

To do exact diagonalization, we assume some orbitals to be frozen (filled background), while allowing a few orbitals to be active. The choice reflects the variational nature of exact diagonalization. In the case of our calculation, we allow the orbitals corresponding to the first valence band of the two valleys to be active. We refer to the valence band that is not filled by an electron (doped hole) in the Hartree-Fock calculation as band 1, and the one that is filled by electron as band 2.

In this constrained Hilbert space, the many-body Hamiltonian is
\begin{widetext}
\begin{equation}
\hat{H}_{d,proj} = \sum_{ij\in \mathrm{active}} \left\{T_{ij} + \sum_{k\in \mathrm{frozen}} \left[V_{ikjk}-V_{ikkj} - \sum_{\gamma \in CNP}(V_{i\gamma j\gamma}-V_{i\gamma \gamma j})\right]\right\}d^{\dagger}_i d^{\dagger}_j
+\frac{1}{2}\sum_{ijkl\in active} V_{ijkl}d^{\dagger}_{i}d^{\dagger}_jd_{l}d_k.
\end{equation}
\end{widetext}
Other terms that take us out of the constrained Hilbert space are left away.
If the frozen orbitals we choose are the same as those filled orbitals  in the Hartree-Fock calculation (frozen=occ), then we can replace
\begin{widetext}
\begin{equation}
T_{ij}+\sum_{k\in \mathrm{frozen}}( V_{ikjk}-V_{ikkj})-\sum_{\gamma \in CNP}(V_{i\gamma j\gamma}-V_{i\gamma \gamma j}) \to \epsilon_i \delta_{ij}.
\end{equation}
\end{widetext}
In this case, we can treat the Hartree-Fock orbital energy as if it is the single-particle energy associated with that orbital. However, in the more general case where the frozen subspace and the filled orbitals in Hartree-Fock calculation are different, e.g., when the frozen subspace is a subset of the later, then
\begin{widetext}
\begin{equation}
\begin{split}
\hat{H}_{d,proj}=\sum_{ij\in \mathrm{active}} \left[\epsilon_{i}\delta_{ij}-\sum_{k\in \mathrm{occ-frozen}} (V_{ikjk}-V_{ikkj})\right]d^{\dagger}_i d^{\dagger}_j
+\frac{1}{2}\sum_{ijkl\in \mathrm{active}} V_{ijkl}d^{\dagger}_{i}d^{\dagger}_jd_{l}d_k.
\end{split}
\end{equation}
\end{widetext}

Our calculation is this latter case, since we dope one hole in our Hartree-Fock calculation. Then \textit{occ-frozen} refers to the valence band that is previously filled in the Hartree-Fock calculation, but now chosen as the active subspace. The magnon gap is defined as flipping the spin/valley of one electron, i.e.,
\begin{widetext}
\begin{equation}
\begin{split}
\Delta_{mag}(n_X)
&= E(N_1=1,N_2=N_{\mathrm{max}}-1,n_X) - E(N_1=0,N_2=N_{\mathrm{max}},n_X)\\
& \quad -\left[E(N_1=1,N_2=N_{\mathrm{max}}-1,0)-E(N_1=0,N_2=N_{\mathrm{max}},0)\right],
\end{split}
\end{equation}
\end{widetext}
where $N_i$ is the particle number in band $i$. For a finite-size study with $N\times N$ momentum grid, $N_{\mathrm{max}}=N^2$. $n_X$ is the exciton density, controlled by the band gap $E_g$. We subtract off the background contribution at $n_X=0$ since we are flipping electrons in the momentum space rather than real space, there are Fock energy changes associated with the flipping process. This does not contribute to the long-range magnetic order.

\section{Wave Packet Calculation\label{wave-packet}}
Here we derive the interaction between a carrier and an exciton. For the sake of concreteness, we take the carrier to be an electron, although taking the carrier to be hole does not change the resulting expression of the exchange interaction.

The starting point of our analysis is the following Hamiltonian, which is just the standard second-quantized Hamiltonian including Coulomb interaction
\begin{equation}
\begin{split}
H = \sum_{i}\epsilon c^{\dagger}_ic_i &+ \frac{1}{2}\sum_{i,j,k,l}\braket{ij|V_c|lk}c^{\dagger}_ic^{\dagger}_jc_kc_l \\
&- \sum_{i,j}\braket{i|V_{single}|j}c^{\dagger}_i c_j .
\end{split}
\end{equation}

In the absence of the Coulomb interaction, the ground state is simply the state with lowest N electron levels occupied. We are concerned with the effect of Coulomb interaction on spin alignment, thus we can safely ignore the kinetic and single-particle potential terms. Additionally, we call the states above the Fermi level "electron states" and states below the "hole states". The creation/annihilation operator of the hole states is the annihilation/creation operator of the electron states ($h^{\dagger}_{i}=e_{i}$). With all these, the Hamiltonian for Coulomb interaction can be separated into three groups, i.e., between electrons, between holes, between electrons and holes.
\begin{align}
H_{ee}&=\frac{1}{2}\sum_{i,j,k,l}\braket{ij|V|kl}e^{\dagger}_ie^{\dagger}_j e_l e_k ,\\
H_{hh}&=\frac{1}{2}\sum_{\alpha,\beta,\gamma,\delta}\braket{\alpha\beta|V|\gamma\delta}h^{\dagger}_\delta h^{\dagger}_\gamma h_\alpha h_\beta ,\\
H_{eh}&=\sum_{i,j,\alpha,\beta}[\braket{i\beta|V|\alpha j}-\braket{i\beta|V|j\alpha}]e^{\dagger}_ih^{\dagger}_{\alpha}h_\beta e_j .
\end{align}

Notice that we have retained only those terms that preserve hole numbers as well as electron numbers. This is because we will be taking matrix elements of the Hamiltonian between states that preserve these quantities. Additionally, when producing $H_{eh}$, we have discarded those $\delta$ terms arsing from the commutation relations, as they only contribute to the kinetic terms. For convenience, we will discuss the scenario where the doped carrier is electron. But the discussion where the doped carrier is hole is completely analogous.

If we take matrix elements of the Hamiltonian between states consisting of one exciton and one electron (e.g. one electron plus one exciton). The following terms will be produced, where $\ket{FS}$ is the Fermi sea state and labels such as $h$ and $e'1$ are compound indices that includes the momentum and spin/valley,
\begin{widetext}
\begin{align}
\begin{split}
&\braket{c_{e'1}c_{e'2}v_{h'}|H_{ee}|c_{e1}c_{e2}v_{h}}\\
&=\frac{1}{2}\sum_{ijkl}V_{ijkl}\delta_{h,h'}\braket{FS|e_{e'2}e_{e'1}e^{\dagger}_{i}e^{\dagger}_je_le_k e^{\dagger}_{e1}e^{\dagger}_{e2}|FS}\\
&=\frac{1}{2}\sum_{ijkl}V_{ijkl}\delta_{h,h'}(\delta_{e'1,i}\delta_{j,e'2}-\delta_{e'1,j}\delta_{i,e'2})(\delta_{l,e2}\delta_{k,e1}-\delta_{k,e2}\delta_{l,e1})\\
&=\delta_{hh'}(V_{e'1e'2,e1e2}-V_{e'1e'2,e2e1})\\
&\sim -\delta_{hh'}V_{e'1e'2,e2e1},
\end{split}
\\[2ex]
\begin{split}
&\braket{c_{e'1}c_{e'2}v_{h'}|H_{eh}|c_{e1}c_{e2}v_{h}}\\
&=\sum_{i,j,\alpha,\beta}(V_{i\beta,\alpha j}-V_{i\beta,j\alpha})\delta_{\beta h}\delta_{h' \alpha}(\delta_{e'2e2}\delta_{i,e'1}\delta_{j,e1}-\delta_{e'2,e1}\delta_{i,e'1}\delta_{j,e2}-\delta_{i,e'2}\delta_{j,e1}\delta_{e'1,e2}+\delta_{e2,j}\delta_{e'1e1}\delta_{i,e'2})\\
&=(V_{e'1h,h'e1}-V_{e'1h,e1h'})\delta_{e'2e2}-(V_{e'1h,h'e2}-V_{e'1h,e2h'})\delta_{e'2e1}\\
&\quad -(V_{e'2h,h'e1}-V_{e'2h,e1h'})\delta_{e'1e2}+(V_{e'2h,h'e2}-V_{e'2h,e2h'})\delta_{e'1e1}\\
&\sim V_{e'1h,e2h'}\delta_{e'2e1}+V_{e'2h,e1h'}\delta_{e'1e2}.
\end{split}
\end{align}
\end{widetext}
Notice that we have used in our derivation the fact that (since V only depends on the magnitude of its argument)
\begin{widetext}
\begin{equation}
\begin{split}
&V_{e'1e'2,e1e2}=\int dr_1 dr_2 \psi^{*}_{e'_1}(r_1)\psi^{*}_{e'_2}(r_2)V(r_1-r_2)\psi^{*}_{e_1}(r_1)\psi^{*}_{e_2}(r_2)=V_{e'2e'1,e2e1}.
\end{split}
\end{equation}
\end{widetext}
And the $\sim$ means omitting direct terms (e.g., $\delta_{hh'}V_{e'1e'2,e1e2}$) and terms that involve the overlapping of wave function on two different layers (e.g., $\delta_{e'2e2}V_{e'1h,h'e1}$).\par

This is a good place to clarify a tricky notation problem: it should also be noticed that when we compute the matrix elements of the Coulomb interaction, we are doing it with respect to the vacuum state rather than the Fermi sea state. Assume $(h'1,h'2,h1,h2)$ are hole labels.
\begin{widetext}
\begin{equation}
\begin{split}
&V_{h'1h'2,h1h2}=\int dr_1 dr_2 \psi^{*}_{h'_1}(r_1)\psi^{*}_{h'_2}(r_2)V(r_1-r_2)\psi_{h_1}(r_1)\psi_{h_2}(r_2).
\end{split}
\end{equation}
\end{widetext}

In the plane wave basis, the Coulomb matrix elements have the following form, where $k_i$ is momentum labels are $\sigma_i$ are spin/valley labels. 
\begin{widetext}
\begin{equation}
\begin{split}
V_{k_1k_2,k_3k_4}&=\frac{1}{A^2}\int dr_1 dr_2 e^{-ik_1r_1}e^{-ik_2r_2}e^{ik_3r_1}e^{ik_4r_2}V(r_1-r_2)\delta_{\sigma_1,\sigma_3}\delta_{\sigma_2,\sigma_4}\\
& =\int d(r_1-r_2)dr_2V(r_1-r_2)e^{i(k3-k1)(r_1-r_2)}e^{i(k_4-k_2+k_3-k_1)r_2}\delta_{\sigma_1,\sigma_3}\delta_{\sigma_2,\sigma_4}\\
&=\frac{V_{l_1l_2}(k_3-k_1)}{A} \delta_{k_1+k_2,k_3+k_4}\delta_{\sigma_1,\sigma_3}\delta_{\sigma_2,\sigma_4},
\end{split}
\end{equation}
\end{widetext}
where $A$ is the system area, $V_{l_i l_j}(k)$ is the Fourier transform  of  Coulomb interaction between layer $i$ and $j$.

The moir\'e trapped electron state has the following expression
\begin{equation}
c^{\dagger}_{j,\sigma}\ket{FS}=\sum_{k_e}e^{-ik_eR_j} W_e(k_e)e^{\dagger}_{k_e,\sigma}\ket{FS}.
\end{equation}
We require the electron to be trapped (up to normalization)
\begin{equation}
\begin{split}
\sum_{k_e} W_e(k_e)e^{ik_e(R_e-R_j)} &\propto e^{-(R_j-R_e)^2/a_e^2}\\
&\Rightarrow W_e(k_e)=\frac{a_e^2}{2}e^{-\frac{1}{4}a_e^2k_e^2}.
\end{split}
\end{equation}
The trapped exciton state has the following expression, where $a^{\dagger}_{k,\sigma_e\sigma_h}$ creates an exciton of COM momentum k. $q_x$ is basically the relative momentum between an electron and a hole.
\begin{widetext}
\begin{align}
\begin{split}
&a^{\dagger}_{j,\sigma_e,\sigma_h}\ket{FS}\\
&=\sum_{k}e^{-ikR_j}W_x(k)a^{\dagger}_{k,\sigma_e\sigma_h}\ket{FS}\\
&=\sum_{k,q_x}e^{-ikR_j}W_x(k)\psi_x(q_x)e^{\dagger}_{q_x+\frac{m_e}{M}k,\sigma_e}h^{\dagger}_{q_x-\frac{m_h}{M}k,\sigma_h}\ket{FS}\\
&=\int \sum_{k,q_x}W_x(k) e^{-ikR_j}dR_e dR_h \psi_x(q_x)
e^{i(q_x+\frac{m_e}{M}k)R_e}e^{-i(q_x-\frac{m_h}{M}k)R_h}e^{\dagger}_{R_e,\sigma_e}h^{\dagger}_{R_h,\sigma_h}\ket{FS}\\
&\propto \int \sum_{k,q_x}e^{-ikR_j}W_x(k) dR_c dR_r \psi_x(q_x) e^{iq_xR_r}e^{ikR_c}
 e^{\dagger}_{R_e (R_c,R_r),\sigma_e}h^{\dagger}_{R_h (R_c,R_r),\sigma_h}\ket{FS}.\\
\end{split}
\end{align}
\end{widetext}
We require the exciton to be trapped (up to normalization)

\begin{equation}
\begin{split}
\sum_k e^{ik(R_c-R_j)}W_x(k) &\propto e^{-(R_c-R_j)^2/{a_x^2}}\\
&\Rightarrow W_x(k)=\frac{a_x^2}{2}e^{-\frac{1}{4}a_x^2k^2}.
\end{split}
\end{equation}
We also require, owing to the exciton being in 1\textit{s} state, that
\begin{equation}
\begin{split}
\sum_{q_x}e^{iq_xR_r}\psi_x(q_x) &\propto e^{-R_r/a_b}\\
&\Rightarrow \psi_x(q_x)=\frac{a_b^2}{(1+a_b^2q_x^2)^{\frac{3}{2}}}.
\end{split}
\end{equation}

We then evaluated the matrix elements in the following basis state (doped electron trapped at $R_i$ and exciton trapped at $R_j$)
\begin{equation}
\begin{split}
&\ket{\Psi_{ij,\sigma\sigma_e\sigma_h}}\\
&=c^{\dagger}_{i,\sigma}a^{\dagger}_{j,\sigma_e\sigma_h}\ket{FS}\\
&=C\sum_{k,q_x,k_e}e^{-ik_eR_i}W_e(k_e)e^{-ikR_j}W_x(k)\\
&\quad \times \psi_x(q_x) e^{\dagger}_{k_e,\sigma}e^{\dagger}_{q_x+\frac{m_e}{M}k,\sigma_e}h^{\dagger}_{q_x-\frac{m_h}{M}k,\sigma_h}\ket{FS},
\end{split}
\end{equation}
with normalization condition 
\begin{equation}
\dfrac{1}{C^2}=\dfrac{1}{C_0^2}-c_1 \delta_{\sigma,\sigma_e},
\end{equation}
in which the expressions
\begin{widetext}
\begin{align}
\frac{1}{C_0^2} &= \frac{A^3}{(2\pi)^6}\frac{\pi^3a_x^2a_e^2a_b^2}{8},\\
c_1 &= \begin{aligned}[t]
&\frac{A^3}{(2\pi)^6}\int dk dq_x dk_e e^{i(q_x+\frac{m_e}{M}k-k_e)(R_i-R_j)}
W_e(k_e)W_e(q_x+\frac{m_e}{M}k)W_x(k)W_x(k_e-q_x+\frac{m_h}{M}k) \\
&\times \psi_x(q_x)\psi_{x}(\frac{m_e}{M}q_x+\frac{m_h}{M}k_e-\frac{\mu}{M}k).
\end{aligned}
\end{align}
\end{widetext}
Defining \(\frac{1}{C_1^2}=\frac{1}{C_0^2}-c_1\), the three terms we need to calculate are
\begin{widetext}
\begin{align}
\begin{split}
&\braket{\Psi_{ij,\sigma'\sigma'_e\sigma'_h}|H_{ee}|\Psi_{ij,\sigma\sigma_e\sigma_h}}\\
&\supseteq -\frac{C_1^2}{A}\sum_{k,q_x,k_e,k',q'_x,k'_e}e^{-i(k_e-k'_e)R_i}e^{-i(k-k')R_j}\psi_x(q_x)\psi^*(q'_x)
 W_e(k_e)W_e^{*}(k_e')W_x(k)W^*_x(k')\\
&\quad \times \delta_{\sigma_h,\sigma_h'}\delta_{\sigma_e,\sigma'} \delta_{\sigma,\sigma_e'}
\delta_{q_x-\frac{m_h}{M}k,q'_x-\frac{m_h}{M}k'}\delta_{k_e+q_x+\frac{m_e}{M}k,k'_e+q'_x+\frac{m_e}{M}k'}
V_{ee}(q_x+\frac{m_e}{M}k-k_e')\\
&=-\frac{C_1^2}{A}\sum_{k,q_x,k_e,k'}e^{i(k-k')(R_i-R_j)}
\psi_x(q_x)\psi^*(q_x-\frac{m_h}{M}(k-k'))
W_e(k_e)W_e^{*}(k_e+k-k')W_x(k)W^*_x(k')\\
&\quad \times \delta_{\sigma_h,\sigma_h'}\delta_{\sigma_e,\sigma'}\delta_{\sigma,\sigma_e'}
V_{ee}(q_x+\frac{m_e}{M}k-k_e-k+k')\\
&\equiv I_{ee},\\
\end{split}
\\[3ex]
\begin{split}
&\braket{\Psi_{ij,\sigma'\sigma'_e\sigma'_h}|H_{eh}|\Psi_{ij,\sigma\sigma_e\sigma_h}}\\
&\supseteq \frac{C_1^2}{A}\sum_{k,q_x,k_e,k',q'_x,k'_e}e^{-i(k_e-k'_e)R_i}e^{-i(k-k')R_j}\psi_x(q_x)\psi^*(q'_x)
 W_e(k_e)W_e^{*}(k_e')W_x(k)W^*_x(k') \\
&\quad \times \delta_{\sigma',\sigma_e}\delta_{k'_e,q_x+\frac{m_e}{M}k}\delta_{\sigma'_e,\sigma}\delta_{\sigma_h,\sigma_h'}
\delta_{q'_x+\frac{m_e}{M}k'+q_x-\frac{m_h}{M}k,k_e+q'_x-\frac{m_h}{M}k'}V_{eh}(q'_x+\frac{m_e}{M}k'-k_e)\\
&=\frac{C_1^2}{A}\sum_{k,q'_x,k'_e,k'}e^{i(k-k')(R_i-R_j)}\psi_x(k'_e-\frac{m_e}{M}k)\psi^*(q'_x)
 W_e(k'_e+k'-k)W_e^{*}(k_e')W_x(k)W^*_x(k')\\
&\quad \times \delta_{\sigma',\sigma_e}\delta_{\sigma'_e,\sigma}\delta_{\sigma_h,\sigma_h'}V_{eh}(q'_x+\frac{m_e}{M}k'-k'-k'_e+k),\\
\end{split}
\end{align}
\end{widetext}
\begin{widetext}
\begin{align}
\begin{split}
&\braket{\Psi_{ij,\sigma'\sigma'_e\sigma'_h}|H_{eh}|\Psi_{ij,\sigma\sigma_e\sigma_h}}\\
&\supseteq \frac{C_1^2}{A}\sum_{k,q_x,k_e,k',q'_x,k'_e}e^{-i(k_e-k'_e)R_i}e^{-i(k-k')R_j}\psi_x(q_x)\psi^*(q'_x) W_e(k_e)W_e^{*}(k_e')W_x(k)W^*_x(k')\\
&\quad \times \delta_{\sigma,\sigma'_e}\delta_{k_e,q'_x+\frac{m_e}{M}k'}
\delta_{\sigma_e,\sigma'}\delta_{\sigma_h,\sigma_h'}
\delta_{k'_e+q_x-\frac{m_h}{M}k,q_x+\frac{m_e}{M}k+q_x'-\frac{m_h}{M}k'}V_{eh}(-q_x-\frac{m_e}{M}k+k'_e)\\
&= \frac{C_1^2}{A}\sum_{k,q_x,k_e,k'}e^{i(k-k')(R_i-R_j)}\psi_x(q_x)\psi^*(k_e-\frac{m_e}{M}k')W_e(k_e)W_e^{*}(k_e+k-k')W_x(k)W^*_x(k')\\
&\quad \times \delta_{\sigma,\sigma'_e}
\delta_{\sigma_e,\sigma'}\delta_{\sigma_h,\sigma_h'}V_{eh}(-q_x-\frac{m_e}{M}k+k_e+k-k')\\
&\equiv I_{eh}.
\end{split}
\end{align}
\end{widetext}
Notice that under the mild assumption that W's and $\psi$'s are all real, then the last two terms are complex conjugate of each other. If we further assume that W's and $\psi$'s are even, then the last two terms are both real. It's also trivial to prove that the term from $H_{ee}$ is real. Thus the effective Hamiltonian takes the following form:
\begin{widetext}
\begin{equation}
\begin{split}
H_{eff} &= \sum_{ij,\sigma,\sigma_e,\sigma_h}(I_{ee}+2I_{eh})c^{\dagger}_{i,\sigma}a^{\dagger}_{j,\sigma_e,\sigma_h}a_{j,\sigma,\sigma_h}c_{i,\sigma_e}
\equiv \sum_{ij,\sigma,\sigma_e,\sigma_h}Jc^{\dagger}_{i,\sigma}a^{\dagger}_{j,\sigma_e,\sigma_h}a_{j,\sigma,\sigma_h}c_{i,\sigma_e}.
\end{split}
\end{equation}
\end{widetext}
We have used $C_1^2$ rather than $C^2$ when calculating the matrix elements. This is because by using $C_1^2$ rather than $C^2$, we have discarded the following term in the effective Hamiltonian
\begin{widetext}
\begin{align}
\begin{split}
H_{dis} &\propto
 \sum_{\sigma,\sigma_h,i,j}c^{\dagger}_{i,\sigma}a^{\dagger}_{j,-\sigma,\sigma_h} a_{j,\sigma,\sigma_h}c_{i,-\sigma}\\
&=\sum_{\sigma,\sigma_h,i,j}c^{\dagger}_{i\sigma}c_{i,-\sigma}c^{\dagger}_{j,-\sigma}c_{j,\sigma}v^{\dagger}_{j,\sigma_h}v_{j,\sigma_h}
 -c^{\dagger}_{i\sigma}c_{i,\sigma}v^{\dagger}_{i,\sigma_h}v_{i,\sigma_h}\delta_{ij}\\
&=\sum_{i,j}N^{hole}_{j}(S^{+}_{i}S^{-}_{j}+S^{-}_{i}S^{+}_{j})
 -\sum_{i}N^{electron}_{i}N^{hole}_{i},
\end{split}
\end{align}
\end{widetext}
$N_i$ counts the particle number at $R_i$. We have used the usual substitution between spin operators and electron creation/annihilation operators
\begin{equation}
\begin{split}
&S^{+}=e^{\dagger}_{\uparrow}e_{\downarrow} ,\\
&S^{-}=e^{\dagger}_{\downarrow}e_{\uparrow} ,\\
&S^{z}=\frac{1}{2}(e^{\dagger}_{\uparrow}e_{\uparrow}-e^{\dagger}_{\downarrow}e_{\downarrow}) .
\end{split}
\end{equation}

Now it's easy to see that the first term in $H_{dis}$ is an in-plane term and it does not contribute to the magnetic order. The second term counts the particle number, and it also does not contribute to the magnetic order.

\bibliography{main}
\end{document}